\documentstyle[twocolumn,epsf,seceq]{jpsj}

\def\runtitle{
Electronic Structure of Stripes in Two-Dimensional Hubbard Model
}

\def\runauthor{
Masanori {\sc Ichioka} and Kazushige {\sc Machida}
}

\title{Electronic Structure of Stripes in Two-Dimensional Hubbard Model}

\author{
Masanori {\sc Ichioka}\footnote{E-mail: oka@mp.okayama-u.ac.jp} and
Kazushige {\sc Machida}\footnote{E-mail: machida@mp.okayama-u.ac.jp}
}

\inst
{Department of Physics, Okayama University, Okayama 700-8530}

\recdate{
June 7, 1999
}

\abst{
Focusing on La$_{2-x}$Sr$_{x}$CuO$_{4}$, we study the stripe 
structure by the self-consistent mean-field theory of the Hubbard 
model. 
By introducing the realistic Fermi surface topology, the SDW-gapped 
insulator is changed to metallic. 
The solitonic features of the stripe structure and the contribution 
of the mid-gap states are presented. 
We consider the band dispersion, the local density of states, the 
spectral weight, and the optical conductivity, associated with the 
solitonic structure. 
These results may provide important information for the 
experimental research of the stripe structure, 
such as the angle-resolved photoemission experiments. 
The ``Fermi surface'' shape is  changed depending on the 
ratio of the incommensurability $\delta$ and the hole density $n_{\rm h}$.  
In real space, only the stripe region is metallic 
when $\delta/n_{\rm h}$ is large. 
}

\kword
{
Stripe structure, high $T_{\rm c}$ cuprates, 
Spectral weight, Hubbard model
}

\begin{document}

%

\sloppy
\maketitle


\fulltext
\halftext

\section{Introduction}
\label{sec:introduction}

The concept of the stripe structure is quite versatile and powerful 
in understanding various spatially modulated orderings ranging from 
insulator to metal, such as spin or charge density waves (SDW or CDW) 
or spin Peierls systems. 
A canonical example of such a stripe description is the 
incommensurate SDW in metallic Cr dilutely doped with Mn or V atoms. 
Various aspects of the SDW under doping are coherently describable in 
terms of the stripe.~\cite{MachidaCr,Fawcett}  
Another canonical example for such description is the spin Peierls 
systems such as in CuGeO$_3$ or other organic materials under a 
magnetic field which effectively acts as changing the electron 
filling.~\cite{Fujita,Fagot,Kiryukhin}
Thus, it is believed that the stripe concept is a useful paradigm for 
low dimensional electron systems with nearly half-fillings.

Anomalous electronic properties in under-doped high $T_{\rm c}$ cuprates 
attract much attention because understanding of the unusual ground 
state may lead us to a clue to a high $T_{\rm c}$ mechanism. 
One of the key features is the magnetic incommensurate (IC) 
structure, i.e., the stripe structure. 
Recently, the static magnetic IC structure is observed by the elastic 
neutron scattering on La$_{2-x}$Sr$_x$CuO$_4$ 
(LSCO).~\cite{Wakimoto,Wakimoto2,Suzuki,TranquadaR}   
It suggests the static stripe structure. 
The static IC structure is also reported on 
La$_{1.6-x}$Nd$_{0.4}$Sr$_x$CuO$_4$ [ref. \citen{TranquadaNd}] 
and La$_{2-x}$Sr$_x$NiO$_{4+y}$ [refs. 
\citen{TranquadaNi} and \citen{Yoshizawa}].\cite{TranquadaR}
In ${\rm Y Ba_2 Cu_3 O_{7-\delta}}$, the IC fluctuations are reported 
by the inelastic neutron experiments.~\cite{Mook,Arai}

Almost a decade ago, one of the authors~\cite{Machida,Kato} predicts 
some of the features of the above static IC spin modulation in 
lightly doped cuprate systems, stressing the charged stripe or the 
solitonic structure as a convenient and universal ``vehicle'' to 
accommodate excess carriers in an otherwise commensurate 
antiferromagnetic (AF) state within a mean-field treatment for a 
simple  two-dimensional Hubbard model. 
Some features of them  were independently found by others at that 
time~\cite{Poilblanc,Zaanen,Schulz} and later.~\cite{Zaanen2,Salkola} 
And they are now confirmed by a more sophisticated method, such as 
DMRG.~\cite{White}
But, since the above  solitonic structure is an insulator in nature 
while high $T_{\rm c}$ cuprates are metallic, it was claimed that the other 
scenario such as frustrating charge segregation is needed.~\cite{Emery}  
However, we can consider the metallic structure in the same framework 
of the selfconsistent mean-field theory, if the realistic Fermi surface 
topology is taken into account by introducing a next nearest neighbor 
site hopping. 
It is still important to study the detailed structure of the stripe 
state in this framework. 
The stripe state in the $d$-$p$ model (i.e., including O-site contribution) 
was studied within the mean-field theory in ref. \citen{Mizokawa}. 

The purpose of this paper is to analyze the stripe structure based on 
the mean field method of the Hubbard model. 
Needless to say, the mean field theory which satisfies variational 
principle for the energy minimum can be justifiable when a long-range 
order exists, such as in the present case. 
Thermal or quantum fluctuations can be treated perturbatively from 
this solution, if it is needed.  
We consider the doping dependence and the effect of the realistic 
Fermi surface topology. 
Our numerical method is described in \S \ref{sec:mean}. 
The profile of the stripe structure and its band structure are 
studied in \S \ref{sec:profile}. 
We also discuss the doping dependence of the incommensurability.
In \S \ref{sec:LDOS}, we study the density of states (DOS) and the 
local density of states (LDOS). 
The latter will be observed by scanning tunneling microscopy (STM) 
experiments. 
In \S \ref{sec:KDOS}, we discuss the spectral weight, i.e., ${\mib 
k}$-resolved DOS. 
Our results are compared with the data of the angle-resolved 
photoemission (ARPES) experiments,~\cite{Ronning,Ino1,Ino2,Ino3} which is 
a powerful method to study Fermi surface topology and 	pseudo-gap 
structure.\cite{Timusk}  
In \S \ref{sec:optical}, we calculate the optical 
conductivity, which detects excitations of the stripe structure, 
and compare with the data.~\cite{Uchida,Ido}.
Our theoretical study shows that these experimental methods can give 
us a lot of useful information of the stripe structure, 
such as the mid gap state or the ${\mib k}$-dependent SDW gap. 
The last section is devoted to summary and discussions. 
Throughout this paper, we set $\hbar = k_{\rm B} =1$ and the lattice 
constant $c=1$. 
Some of our results were briefly reported in ref. 
\citen{MachidaL}. 

\section{Selfconsistent Mean-Field Theory}
\label{sec:mean}

We start out with the standard Hubbard model on a two-dimensional 
square lattice and introduce the mean field 
\begin{equation}
\langle n_{i,\sigma}\rangle=\frac{1}{2}(n_i+\sigma M_i )
\label{eq:2.1}
\end{equation}
at $i$-site, where $n_i$ ($M_i)$ is the charge (spin) density and 
$i=(i_x,i_y)$. 
Thus, the one-body Hamiltonian is given by  
\begin{eqnarray}
{\cal H}&=&
-\sum_{i,j,\sigma} t_{i,j}a^{\dagger}_{i,\sigma} a_{j,\sigma}
\nonumber
\\
&&
+U\sum_{i,\sigma}\langle n_{i,-\sigma}\rangle n_{i,\sigma} 
-U \sum_i \langle n_{i,\downarrow}\rangle \langle n_{i,\uparrow} 
\rangle.  
\label{eq:2.2}
\end{eqnarray}
where $n_{i,\sigma}=a^{\dagger}_{i,\sigma} a_{i,\sigma}$ ($\sigma=1$ 
for $\uparrow$, $-1$ for $\downarrow$) and $a^{\dagger}_{i,\sigma}$ 
($a_{i,\sigma}$) is a creation (annihilation) operator. 
For the nearest neighbor (NN) pairs $(i,j)$, $t_{i,j}=t$. 
For the next NN pairs situated on a diagonal position in a square 
lattice, $t_{i,j}=t'$. 
For the third neighbor pairs, which are situated along the NN bond 
direction, $t_{i,j}=t''$.
For the stripe structure with the $N$ site periodicity, 
\begin{equation}
\langle n_{i,\sigma}\rangle =\sum_{0 \le l < N} 
{\rm e}^{{\rm i} l{\mibs Q}\cdot {\mibs r}_i} \langle n_{l{\mibs 
Q},\sigma}\rangle, 
\label{eq:2.3}
\end{equation}
where ${\mib r}_i$ is the position of $i$-site. 
The modulation vector is given by ${\mib Q}= {\mib Q}_{\rm D} =2 \pi 
(\frac{1}{2}-\delta,\frac{1}{2}-\delta)$ for diagonal incommensurate 
(DIC) state, ${\mib Q}= {\mib Q}_{\rm V} = 2 \pi 
(\frac{1}{2},\frac{1}{2}-\delta)$ for vertical incommensurate (VIC) 
state, and ${\mib Q}= {\mib Q}_{\rm C}= 2 \pi 
(\frac{1}{2},\frac{1}{2})$ for commensurate antiferromagnetic (C-AF) 
state. 
We set the incommensurability $\delta=M/N$ (rational fraction) with 
even integer $N$. 
We consider eq. (\ref{eq:2.2}) in the reciprocal space. 
There, ${\mib k}$ is coupled to ${\mib k}+l{\mib Q}$ in the term $U 
\sum_{{\mibs k},l,\sigma} \langle n_{l{\mibs Q},-\sigma}\rangle 
a^\dagger_{{\mibs k}+l{\mibs Q},\sigma} a_{{\mibs k},\sigma}$, where 
$a_{{\mibs k},\sigma}$ is the Fourier component of $a_{i,\sigma}$. 
We write ${\mib k}={\mib k}_0+m{\mib Q}$ ($m=0,1,\cdots,N-1$), where 
${\mib k}_0$ is within the reduced Brillouin zone of the size $(2 
\pi)^2/N$. 
Then, eq. (\ref{eq:2.2}) is reduced to 
\begin{eqnarray}
{\cal H} &=&{\cal H}_1+{\cal H}_2
\label{eq:2.4}
\\
{\cal H}_1 &=& 
\sum_{{\mibs k}_0,\sigma,m} \biggl\{ 
\epsilon({\mib k}_0 + m{\mib Q}) 
a^\dagger_{{\mibs k}_0 + m{\mibs Q},\sigma} 
a_{{\mibs k}_0 + m{\mibs Q},\sigma}
\nonumber 
\\ && 
+U\sum_{0<l<N} \langle n_{l{\mibs Q},-\sigma} \rangle
a^\dagger_{{\mibs k}_0 + (m+l){\mibs Q},\sigma} 
a_{{\mibs k}_0 + m{\mibs Q},\sigma} \biggr\}
\nonumber
\\
&=&
\sum_{{\mibs k}_0,\sigma,\alpha}E_{{\mibs k}_0,\sigma,\alpha} 
\gamma^\dagger_{{\mibs k}_0,\sigma,\alpha} \gamma_{{\mibs 
k}_0,\sigma,\alpha}, 
\label{eq:2.5}
\\
{\cal H}_2 &=&
U\sum_{{\mibs k}_0,\sigma,m} \bar{n}_{-\sigma}
a^\dagger_{{\mibs k}_0 + m{\mibs Q},\sigma} a_{{\mibs k}_0 + m{\mibs 
Q},\sigma}
\nonumber
\\ &&
-U N_k \sum_{0 \le l<N} 
\langle n_{l{\mibs Q},\uparrow} \rangle
\langle n_{l{\mibs Q},\downarrow} \rangle, 
\label{eq:2.6}
\end{eqnarray}
where $\bar{n}_\sigma=\langle n_{l=0,\sigma}\rangle$ is the average 
number of $\sigma$-spin electron per site, $N_k=\sum_{i}1=\sum_{{\mibs 
k}_0,m}1$, and $\epsilon({\mib k})=\epsilon_0({\mib k})+\epsilon_1({\mib 
k})$ with $\epsilon_0({\mib k})=-2t(\cos k_x + \cos k_y)$ and 
$\epsilon_1({\mib k})=-4t' \cos k_x \cos k_y -2t''(\cos 2 k_x + \cos 2 
k_y)$. 
In eq. (\ref{eq:2.5}), we diagonalize $N \times N$ matrix ${\cal 
H}_1$ by using the unitary transformation 
\begin{equation}
a_{{\mibs k}_0 + m{\mibs Q},\sigma} 
=\sum_{\alpha} u_{{\mibs k}_0,\sigma,\alpha,m} \gamma_{{\mibs 
k}_0,\sigma,\alpha}, 
\label{eq:2.7}
\end{equation}
where $\alpha(=1,2,\cdots,N)$ is the label of the eigen-energy 
$E_{{\mibs k}_0,\sigma,\alpha}$ and the wave function $u_{{\mibs 
k}_0,\sigma,\alpha,m}$. 

The self-consistent condition is given by 
\begin{eqnarray}
\langle n_{l{\mibs Q},\sigma} \rangle 
&=& 
N_k^{-1} \sum_{{\mibs k}_0,m} \langle a^\dagger_{{\mibs k}_0 + m{\mibs 
Q},\sigma}
a_{{\mibs k}_0 + (m+l){\mibs Q},\sigma} \rangle 
\nonumber \\
&=&
N_k^{-1} \sum_{{\mibs k}_0,m,\alpha} 
u^\ast_{{\mibs k}_0,\sigma,\alpha,m} 
u_{{\mibs k}_0,\sigma,\alpha,m+l} f(E_{{\mibs k}_0,\sigma,\alpha}). 
\nonumber \\
\label{eq:2.8}
\end{eqnarray}
Here, we use $\langle \gamma^\dagger_{{\mibs k}_0,\sigma,\alpha} 
\gamma_{{\mibs k}_0,\sigma,\alpha'} \rangle 
=\delta_{\alpha,\alpha'}f(E_{{\mibs k}_0,\sigma,\alpha}) $ with the 
Fermi distribution function $f(E)=({\rm e}^{(E-\mu)/T}+1)^{-1}$. 

In our calculation, the eigen-energy and the wave functions are 
obtained by eq. (\ref{eq:2.5}) under the given $\langle n_{l{\mibs 
Q},\sigma} \rangle$. 
The chemical potential $\mu$ is determined by the condition 
$\bar{n}_\uparrow=\bar{n}_\downarrow=\frac{1}{2}(1-n_{\rm h})$ for $1-n_{\rm h}$ 
filling. 
Then, we calculate $\langle n_{l{\mibs Q},\sigma} \rangle$ by eq. 
(\ref{eq:2.8}), and use it in the next step calculation of eq. 
(\ref{eq:2.5}). 
This iteration procedure is repeated until a sufficient 
self-consistent solution is obtained. 
Among the possible stripe structures  (C-AF, DIC and VIC with various 
incommensurability $\delta$), the stable structure is determined by 
the total energy ($\langle \cal{H} \rangle$) minimum at zero 
temperature.  
We mainly consider the case $t'=-\frac{1}{6}t$ and $t''=0$. 
This set of the parameters is proposed for LSCO to reproduce the 
Fermi surface obtained by the band 
calculation.\cite{Tanamoto,Mattheiss,Xu}

\section{Spatial Profile and Band Structure}
\label{sec:profile}

For $t'=0$, since the Fermi surface is almost a square shape near 
half-filling ($n_{\rm h} \sim 0$), the complete nesting occurs,  and one 
dimensional (1D) picture may be applicable.~\cite{Machida} 
By introducing the effect of $t'$, the Fermi surface deviates from 
the square shape. 
At first, we consider the $t'$-effect on the C-AF state. 
The energy dispersion is split into the upper conduction band  
and the lower valence band, and it is given by 
\begin{eqnarray}
E_{\pm}({\mib k})
&=& 
\frac{1}{2}[ \epsilon({\mib k})+\epsilon({\mib k}+{\mib Q}_{\rm C}) 
\nonumber \\ &&
\pm \{ (\epsilon({\mib k})-\epsilon({\mib k}+{\mib Q}_{\rm C}))^2 
+4 |\Delta_{\rm AF}|^2 \}^{1/2} ] 
\nonumber \\ 
&=&
\epsilon_1({\mib k}) \pm \{ \epsilon_0({\mib k})^2+|\Delta_{\rm AF}|^2 
\}^{1/2}, 
\label{eq:3.1}
\end{eqnarray}
with the AF gap $\Delta_{\rm AF}$. 
The AF gap opens at ${\mib k}_Q=(k_x,\pm \pi-k_x)$ and $(k_x,\pm 
\pi+k_x)$ by the nesting of ${\mib Q}_{\rm C}$. 
Then, the lower and upper edges of the AF gap are written as 
$E_{\pm}({\mib k}_Q)=\epsilon_1({\mib k}_Q) \pm \Delta_{\rm AF}$. 
For $t'<0$, $E_{\pm}({\mib k}_Q)$ is small near $(1,0)$ and $(0,1)$ in 
reciprocal lattice units compared with near 
$(\frac{1}{2},\frac{1}{2})$, 
while the gap $E_+({\mib k}_Q)-E_-({\mib k}_Q)=2 \Delta_{\rm AF}$ does 
not depend on the ${\mib k}$-point. 
This dispersive AF gap structure is in fact observed by ARPES on some 
parent compounds of High $T_{\rm c}$ superconductors such as a half-filled 
Mott insulator ${\rm Ca_2CuO_2Cl_2}$.~\cite{Ronning} 

\begin{figure}[t]
\begin{center}
\leavevmode
\epsfbox{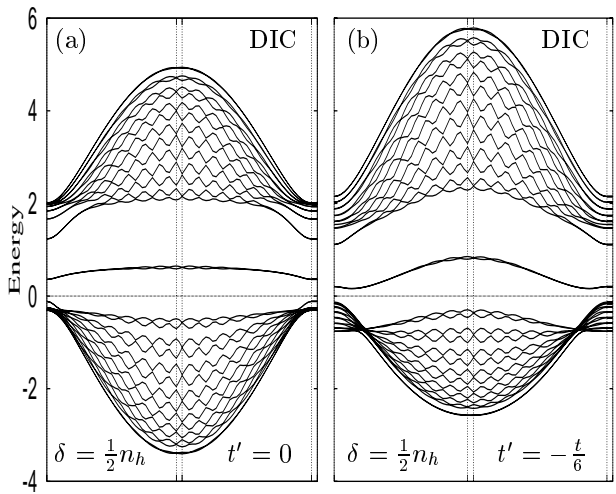}
\end{center}
\caption{
Dispersion relations for DIC along the perimeter of the reduced 
Brillouin zone for $n_{\rm h}=1/16=0.0625$, $U=3.6t$, $\delta/n_{\rm h}=1/2$. 
$t'/t=0$ (a) and $-1/6$ (b). 
The Fermi energy $E_{\rm F}$ is at zero. 
Energy is scaled by $t$. 
}
\label{fig:disp-d}
\end{figure}

\begin{figure}[t]
\begin{center}
\leavevmode
\epsfbox{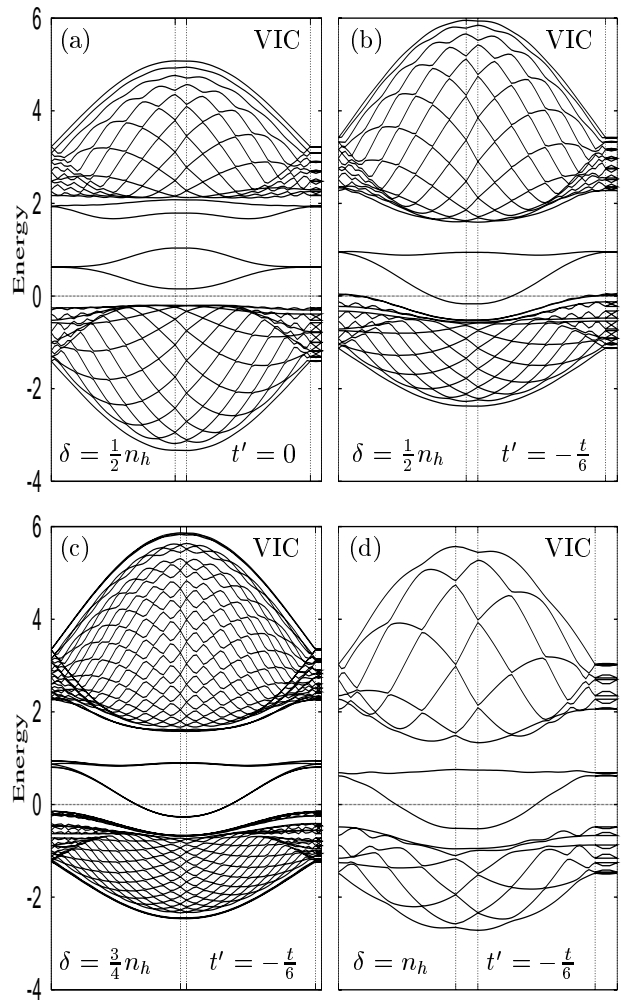}
\end{center}
\caption{
Dispersion relations for VIC along the perimeter of the reduced 
Brillouin zone for $n_{\rm h}=1/16=0.0625$, $U=3.6t$. 
(a) $t'/t=0$ and $\delta/n_{\rm h}=1/2$. 
(b) $t'/t=-1/6$ and $\delta/n_{\rm h}=1/2$.
(c) $t'/t=-1/6$ and $\delta/n_{\rm h}=3/4$. 
(d) $t'/t=-1/6$ and $\delta/n_{\rm h}=1$.
The Fermi energy $E_{\rm F}$ is at zero.
Energy is scaled by $t$. 
}
\label{fig:disp-v}
\end{figure}

Next, we consider the $t'$-effect on the IC state for finite $n_{\rm h}$. 
We also find the dispersive gap structure in the SDW gap.  
We show the energy dispersion within the reduced Brillouin zone in 
Fig. \ref{fig:disp-d} for DIC and in Fig. \ref{fig:disp-v} for VIC in 
the case $n_{\rm h}=1/16=0.0625$. 
There, a large SDW (or AF) gap opens between the conduction band and 
valence band, and there are some mid-gap bands in the large SDW gap. 
In the figure, unit vectors of the reduced Brillouin zone are given 
by ${\mib u}_1=(0,2\pi)$, ${\mib u}_2=(2\pi /N,2\pi /N)$ for DIC, and 
${\mib u}_1=(\pi,\pi - 2\pi M/N)$, ${\mib u}_2=(0,4\pi /N)$ for VIC. 
The horizontal axes in Figs. \ref{fig:disp-d} and \ref{fig:disp-v} 
are along the path $\frac{1}{2}({\mib u}_1 +{\mib u}_2)$ - 
$\frac{1}{2}{\mib u}_2 $ - ${\mib 0}$ - $\frac{1}{2}{\mib u}_1 $ - 
$\frac{1}{2}({\mib u}_1 +{\mib u}_2)$. 


In the stripe structure with $N/M$ periodicity (i.e., $M$ periods in 
$N$ sites), the size of the Brillouin zone is reduced to $1/N$, and 
the energy dispersion is split to $N$ bands as seen in Figs. 
\ref{fig:disp-d} and \ref{fig:disp-v}.
If it were half-filling, lower $N/2$ bands would be occupied by 
electrons. 
In the case of finite doping $n_{\rm h}=M/(N/2)$, the induced holes enter M 
bands among $N/2$ bands. 
Then, an SDW gap opens above the fully occupied $N/2-M$  bands, 
and $2M$ bands above the filled bands form a mid-gap state. 
Since the gapped state minimizes the total energy, the stripe 
structure with $N/M$ periodicity is stable in an insulator state. 
Then, the incommensurability is exactly given by 
$\delta=M/N=\frac{1}{2}n_{\rm h}$ in this insulator state. 
This relation between the incommensurability $\delta$ and the hole 
density $n_{\rm h}$ is confirmed by our energy estimate at $T=0$ in the 
insulator state, even for finite $t'$. 
This is a generic statement which might be important to interpret the 
neutron diffraction 
experiments.\cite{Wakimoto,Wakimoto2,Suzuki,TranquadaR,TranquadaNd,TranquadaNi,Yoshizawa} 

For $t'=0$, the lower and upper edges of the SDW gap are almost flat 
[Figs. \ref{fig:disp-d}(a) and  \ref{fig:disp-v}(a)]. 
But, for finite $t'$, the gap edges become dispersive [Figs. 
\ref{fig:disp-d}(b) and \ref{fig:disp-v}(b)-(d)]. 
With increasing $|t'|$,  both the valence band and the mid-gap state 
touch the Fermi energy $E_{\rm F}$ in the VIC case with 
$\delta=\frac{1}{2}n_{\rm h}$. 
And the stripe state becomes metallic [Fig. \ref{fig:disp-v}(b)]. 
We set $E_{\rm F}=0$ in the figures. 
For smaller $U$,  it becomes easily metallic since the SDW gap is 
small.  

The total energy of DIC is lower than that of VIC for higher $U$ and 
smaller $n_{\rm h}$.\cite{Kato,Zaanen,MachidaL} 
For example, the DIC is stable for $n_{\rm h} < n_{\rm c} \sim 0.1$ 
in the case $U=3.6 t$ and $t'=-\frac{1}{6}t$. 
With raising (lowering) $U$, $n_{\rm c}$ increases (decreases). 
When $|t'|$ is larger, $n_{\rm c}$ decreases.
The DIC state is difficult to become metallic since the dispersion of 
the mid-gap state is flat as shown in Fig. \ref{fig:disp-d}. 
While the metallic DIC state can be  realized  for larger $n_{\rm h}$ and 
smaller $U$, the VIC state is stable in these parameter regions.
In the parameter region of the stable DIC, it is an insulator. 
Therefore, it is unlikely that the DIC is realized to be metallic 
upon doping. 
This is a generic statement within the present framework. 

On the other hand, for the electron doping case ($n_{\rm h}<0$), 
the C-AF state has lower energy than those of VIC and DIC 
for small $|n_{\rm h}|$ and lower $U$. 
And the VIC is stable for large $|n_{\rm h}|$ and higher $U$. 
The DIC state does not appear for $n_{\rm h}<0$. 
The stable VIC's parameter region becomes narrower with increasing $|t'|$.  
The VIC for $n_{\rm h}<0$ is an insulator, because the top of the mid-gap band 
is flat as seen in Fig. \ref{fig:disp-v}(b). 
$E_{\rm F}$ is located in the gap between the mid-gap band and the upper conduction band. 

\begin{figure}[t]
\begin{center}
\leavevmode
\epsfbox{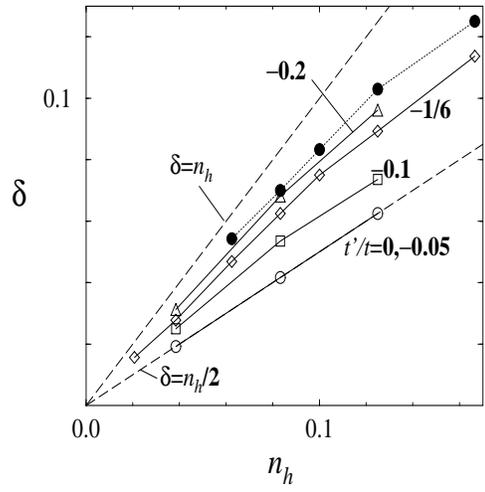}
\end{center}
\caption{
The $n_{\rm h}$ dependence of the incommensurability $\delta$  to minimize 
the total energy for VIC. 
Open symbols with solid lines are for $t'/t=0$, $-0.05$ (insulator), 
$-0.1$, $-1/6$. $-0.2$ (metal). 
Filled circles with a dotted line are for $t'=-0.12t$ and 
$t''=0.08t$. 
The lines are guide for the eye. 
In metallic case, $\delta$ is enhanced with increasing $|t'|$. 
We also show lines for $\delta=n_{\rm h}$ and $\delta=n_{\rm h}/2$ (dashed lines).
}
\label{fig:IC}
\end{figure}

\begin{figure}[t]
\begin{center}
\leavevmode
\epsfbox{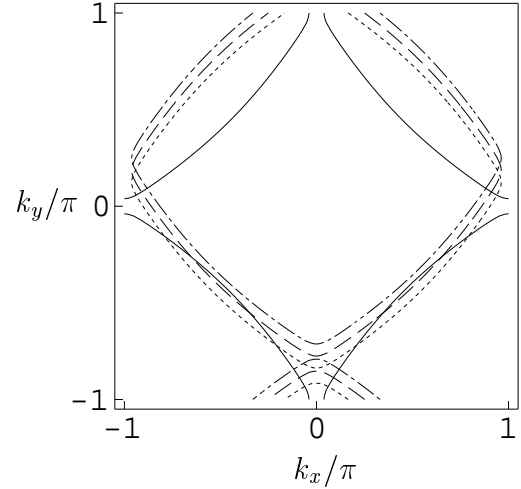}
\end{center}
\caption{
Nesting features: the original Fermi surface of 
$\epsilon({\mib k})$ [solid lines] and ${\mib Q}_{\rm V}$-shifted ones 
for $\delta/n_{\rm h}$=1/2 [dotted lines], 3/4 [dashed lines] and 1 
[dash-dotted lines]. 
$n_{\rm h}=0.125$, $t'/t=-1/6$, $t''=0$. 
The $2 \pi \times 2 \pi$ Brillouin zone is presented. 
The nesting is seen to be better for $\delta/n_{\rm h} \sim 3/4$ than $1/2$ 
or $1$. 
}
\label{fig:nesting}
\end{figure}

Once the metallic state is realized for $n_{\rm h}>0$, the relation 
$\delta=\frac{1}{2}n_{\rm h}$, which is valid for the insulating case 
due to the  energy gain by the SDW gap, is not obeyed.  
Figure \ref{fig:IC} shows the $n_{\rm h}$ dependence of $\delta$ for VIC by 
our estimate of the total energy minimum. 
There, $t'/t$ is changed from 0 to $-0.2$, and $t''=0$. 
It is found that $\delta$ is fixed to be $\frac{1}{2}n_{\rm h} $ in 
insulator state ($t'/t=0$, $-0.05$).  
But, in metallic state of VIC, $\delta$ becomes larger than the 
$\frac{1}{2}n_{\rm h} $-line, 
and approach the $n_{\rm h}$-line with increasing $|t'|$.  
The $n_{\rm h}$-dependence is almost linear at small $n_{\rm h}$, and slightly 
suppressed at large $n_{\rm h}$. 
The $n_{\rm h}$ dependence of $\delta$  dose not depend on $U$ in the 
metallic state for $3 \le U/t \le 4$.  
It only depends on $t'$, i.e., the shape of the Fermi surface. 
It may be related to the nesting of ${\mib Q}$ vector. 
Figure \ref{fig:nesting} shows the Fermi surface of $\epsilon({\mib 
k})$ for $n_{\rm h} =0.125$ and the ${\mib Q}_{\rm V}$-shifted Fermi 
surface. 
The nesting seems to be better in the case $\delta\sim\frac{3}{4}n_{\rm h}$  
than in the case $\delta\sim\frac{1}{2}n_{\rm h}$.  
It is the reason why $\delta$ is larger than $\frac{1}{2}n_{\rm h}$. 
With increasing $n_{\rm h}$ further, the parameter $\delta/n_{\rm h}$ for the 
best nesting is shifted to smaller. 
Then, the saturation of $\delta$ occurs at larger $n_{\rm h}$ as seen from 
Fig. \ref{fig:IC}. 
We also consider the case $t'=-0.12t$ and $t''=0.08t$. 
These parameters are proposed to reproduce the Fermi surface obtained 
by the ARPES experiments.~\cite{Ino2,Ino3,Tohyama}  
The obtained $\delta$ is near the $n_{\rm h}$-line, as shown by filled 
circles in Fig. \ref{fig:IC}. 
We note that the C-AF state has lower energy than VIC and DIC for small 
$n_{\rm h}$ in the case $U$ is small and $\delta /n_{\rm h}$ approaches 1.  
As a result, $\delta=\frac{1}{2} n_{\rm h} $ for insulator DIC and VIC. 
And for metallic VIC, $\delta$ approaches $\delta=n_{\rm h}$ with 
increasing $|t'|$ or $|t''|$. 
As for the DIC case, $\delta =\frac{1}{2} n_{\rm h} $ in the insulator 
state. 
In the metallic DIC, which is difficult to be stable, $\delta$ is 
suppressed than $\frac{1}{2} n_{\rm h} $-line. 


\begin{figure}[t]
\begin{center}
\leavevmode
\epsfbox{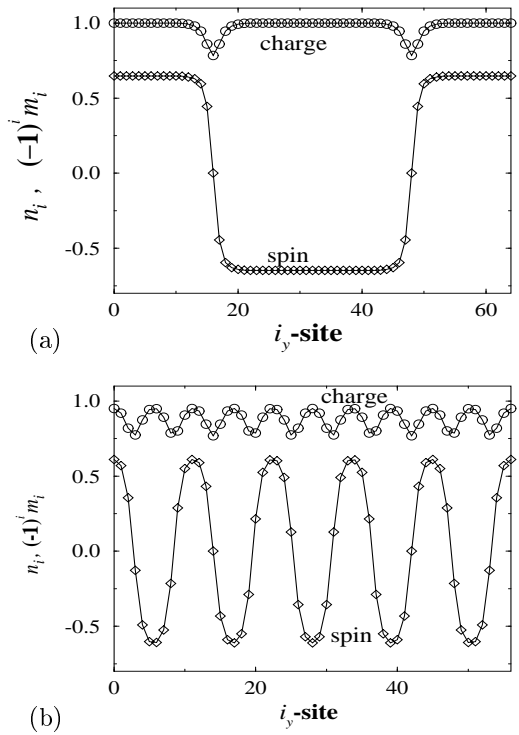}
\end{center}
\caption{
The spatial profile of the charge density $n_i$ and the magnetization 
$(-1)^{i_x+i_y} m_i$ along the $y$-direction in VIC. 
$U=3.6t$, $t'/t=-1/6$. 
(a) $n_{\rm h}=1/48=0.021$, $\delta/n_{\rm h}=3/4$. 
(b) $n_{\rm h}=1/8=0.125$, $\delta/n_{\rm h}=5/7$. 
}
\label{fig:profile}
\end{figure}

Some of our results are consistent with the experimental results. 
The elastic and inelastic neutron scattering experiments suggest that 
the insulator La$_{2-x}$Sr$_x$NiO$_{4+y}$ is DIC with $\delta \sim 
\frac{1}{2}n_{\rm h}$.\cite{TranquadaR,TranquadaNi,Yoshizawa} 
The metallic (or superconducting) LSCO for $x>0.06$ and 
La$_{1.6-x}$Nd$_{0.4}$Sr$_x$CuO$_4$ are 
VIC.~\cite{Wakimoto,Wakimoto2,Suzuki,TranquadaR,TranquadaNd}  
There, it is suggested that $\delta  \sim n_{\rm h}$ for small $n_{\rm h}$. 
The saturation of $\delta$ occurs for large $n_{\rm h}$, where the 
inter-stripe distance is close to the width of each stripe as shown 
later in Fig. \ref{fig:profile}(b). 
The upper limit of $\delta$ seems to be determined by the width of 
each stripe, beyond which $\delta$ saturates. 
The LSCO exhibits the transition from metallic to insulator at $x 
\sim 0.05$.
With this transition, VIC with $\delta \sim x$ ($\sim n_{\rm h}$) is 
changed to DIC with $\delta \sim  \frac{1}{2} x$ [in the notation of 
ref. \citen{Wakimoto}, ${\mib Q}_{\rm 
D}=2\pi(\frac{1}{2}-\frac{\epsilon}{2}, 
\frac{1}{2}-\frac{\epsilon}{2})$ with $\epsilon \sim n_{\rm h}$].  
Reference \citen{Wakimoto} reported that $\epsilon=0.06 \pm 0.005$ at 
$x=0.05$ in DIC. 
Also in our model, VIC is changed to DIC at critical $n_{\rm h}$ with 
decreasing $n_{\rm h}$ for appropriate $U$.
The DIC is insulator with $\delta = \frac{1}{2}n_{\rm h}$, and VIC can be 
metallic. 
Our model shows that $\delta$ becomes larger than 
$\frac{1}{2}n_{\rm h}$ in the metallic VIC case.  
But, it fails to give the 
relation exactly $\delta = n_{\rm h}$. 
To improve this point, we need to carefully adjust the parameters of the 
Fermi surface topology.~\cite{Kuroki} 
On this problem, the effect of the long range Coulomb interaction 
was also suggested.~\cite{Seibold} 
There is a possibility that the VIC state with $\delta = n_{\rm h}$ 
becomes more stable by adding the long range Coulomb interaction to  
the Fermi surface topology effect. 

To see the effect of the $\delta$ shift on the gap structure at fixed 
$n_{\rm h}$, we show the energy dispersion  for 
$\delta = \frac{3}{4}n_{\rm h}$ (c) [which gives the total energy minimum] 
and  $\delta = n_{\rm h}$ (d) in Fig. \ref{fig:disp-v}. 
They show qualitatively the same band structure as in the case 
$\delta = \frac{1}{2}n_{\rm h}$ [Fig. \ref{fig:disp-v}(b)]. 
But, the position of $E_{\rm F}$ is changed depending on $\delta$. 
With increasing $\delta$ at fixed $n_{\rm h}$ (i.e., effectively decreasing 
$n_{\rm h}$ compared with $\delta$), $E_{\rm F}$ is shifted toward higher 
energy in the band structure. 
Both the valence band and the mid-gap state cross $E_{\rm F}$ for  
$\delta=\frac{1}{2}n_{\rm h}$. 
For $\delta=\frac{3}{4}n_{\rm h}$ and $\delta = n_{\rm h}$, only the mid-gap 
state crosses $E_{\rm F}$, and the valence band does not appear 
at $E_{\rm F}$. 
In the following, we consider the optimized $\delta$ case shown in Fig. 
\ref{fig:IC} ($\delta \sim \frac{3}{4}n_{\rm h}$ for $t'=-\frac{1}{6}t$). 
But the result is not qualitatively changed for other $\delta$. 
To know other $\delta$ cases, it is enough to shift $E_{\rm F}$ in 
the following results, keeping the overall gap structure unchanged.

\begin{figure}[t]
\begin{center}
\leavevmode
\epsfbox{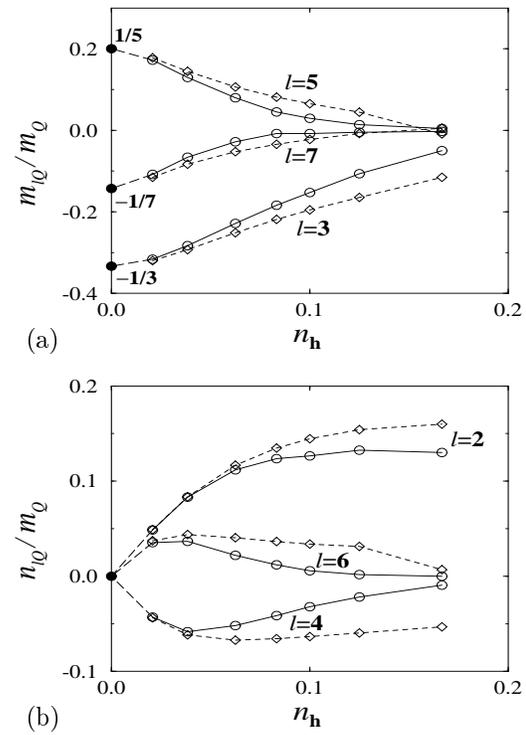}
\end{center}
\caption{
The $n_{\rm h}$-dependence of the higher harmonics of the Fourier 
components, $m_{l{\mibs Q}}$ (a) and $n_{l{\mibs Q}}$ (b) for VIC 
($\circ$) and DIC ($\diamond$). 
We plot $m_{l{\mibs Q}}/m_{{\mibs Q}}$ ($l=3,5,7$) and $n_{l{\mibs 
Q}}/m_{{\mibs Q}}$ ($l=2,4,6$), respectively. 
Lines are guide for the eye, and $\bullet$ shows the extrapolation 
for $n_{\rm h} \rightarrow 0$. 
$U=3.6t$, $t'/t=-1/6$. 
We use the optimized value of $\delta$ given in Fig. 
\protect{\ref{fig:IC}} 
}
\label{fig:Fc}
\end{figure}

Next, we discuss the spatial profile of the stripe structure. 
In Fig. \ref{fig:profile}, we show the spatial profiles of the charge 
density $n_i$ and the magnetization $(-1)^{i_x+i_y} m_i$ along the 
$y$-direction. 
The excess carriers are located in the stripe region, and form CDW. 
The width of the charged stripe is about 3 sites in this case. 
The AF changes its phase by $\pi$ when it crosses the stripe region. 
The envelope of $m_i$ looks like sinusoidal wave for large $n_{\rm h}$,  
but it looks like square wave for small $n_{\rm h}$. 
To clearly see this profile change, we  consider the Fourier 
components $n_{l{\mibs Q}}= n_{l{\mibs Q},\uparrow} + n_{l{\mibs 
Q},\downarrow}$ and $m_{l{\mibs Q}}= n_{l{\mibs Q},\uparrow} - n_{l{\mibs 
Q},\downarrow}$. 
In Fig. \ref{fig:Fc}, we plot the $n_{\rm h}$-dependence of $m_{l{\mibs 
Q}}/m_{{\mibs Q}}$ ($l=3,5,7$) and $n_{l{\mibs Q}}/m_{{\mibs Q}}$ 
($l=2,4,6$). 
For large $n_{\rm h}$, $m_{l{\mibs Q}}/m_{{\mibs Q}} \rightarrow 0 $ ($l \ge 
3$). 
It means that the fundamental component $m_{{\mibs Q}}$ is dominant and 
that $m_i$ varies as a simple sinusoidal wave. 
On the other hand, with decreasing $n_{\rm h}$, $m_{l{\mibs Q}}/m_{{\mibs Q}} 
\rightarrow (-1)^{(l-1)/2}/l $ ($l \ge 3$: odd integer). 
It is the Fourier component of the square wave. 
The continuous change from the sinusoidal wave to the square wave 
with decreasing $n_{\rm h}$ is reminiscent of the Jacobi elliptic function 
$sn(x,k)$, which was the form in 1D CDW or SDW systems near the 
half-filling.\cite{Machida}

\begin{figure}[t]
\begin{center}
\leavevmode
\epsfbox{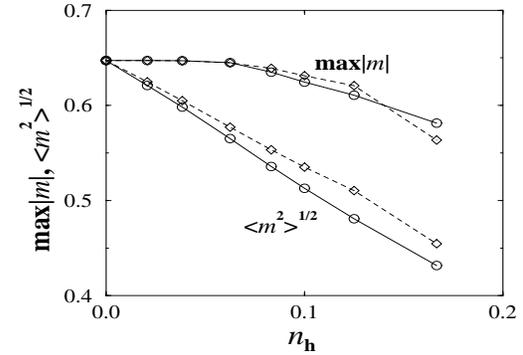}
\end{center}
\caption{
The $n_{\rm h}$-dependence of the maximum (${\rm max}|m_i|$) and the 
variance ($\langle m_i^2 \rangle^{1/2}$) of the magnetization in VIC 
($\circ$) and DIC ($\diamond$).
$U=3.6t$, $t'/t=-1/6$.  
}
\label{fig:mag}
\end{figure}
\begin{figure}[t]
\begin{center}
\leavevmode
\epsfbox{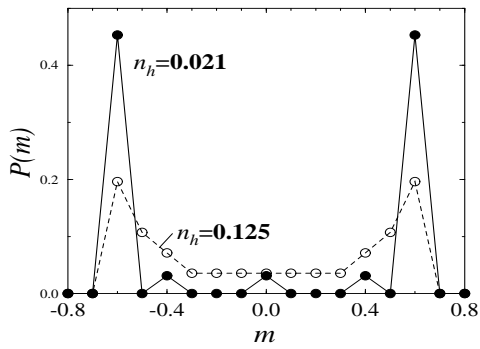}
\end{center}
\caption{
The distribution function of the magnetic moment, $P(m)$, for 
$n_{\rm h}=1/48=0.021$ ($\bullet$) and for $n_{\rm h}=1/8=0.125$ ($\circ$) in 
VIC, which correspond to the cases of Figs. 
\protect{\ref{fig:profile}}(a) and \protect{\ref{fig:profile}}(b), 
respectively. 
Lines are guides for the eye. 
}
\label{fig:mag-dis}
\end{figure}
\begin{figure}[t]
\begin{center}
\leavevmode
\epsfbox{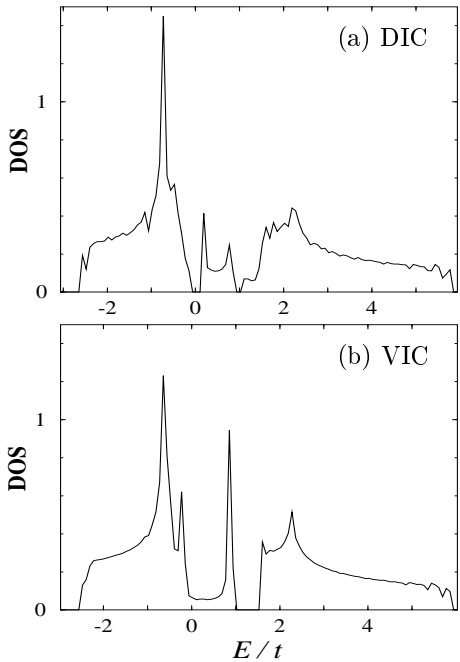}
\end{center}
\caption{
The density of states $N(E)$. 
$U=3.6t$, $t'/t=-1/6$, $n_{\rm h}=1/16=0.0625$. 
(a) The insulator DIC case of Fig. \protect{\ref{fig:disp-d}}(b). 
$\delta/n_{\rm h}=1/2$.  
(b) The metallic VIC case of Fig. \protect{\ref{fig:disp-v}}(c).  
$\delta/n_{\rm h}=3/4$.  
The Fermi energy $E_{\rm F}$ is at 0.
}
\label{fig:DOS}
\end{figure}
\begin{figure}[t]
\begin{center}
\leavevmode
\epsfbox{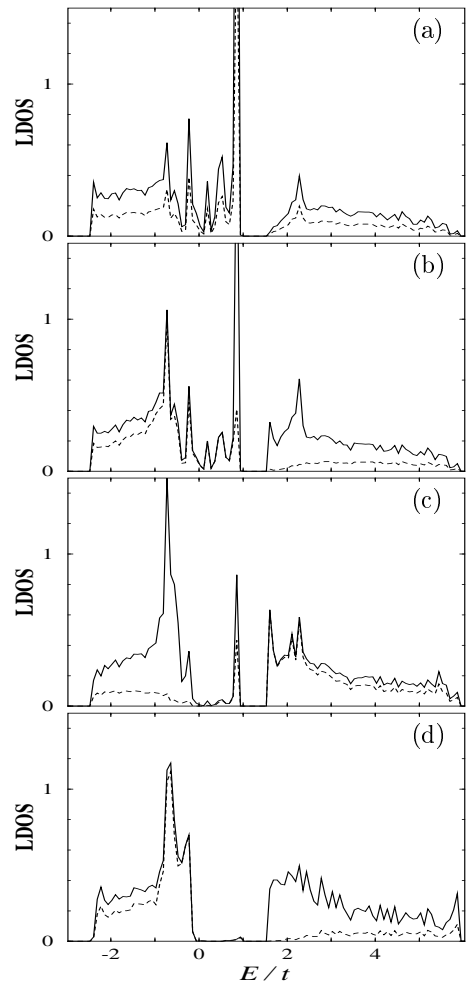}
\end{center}
\caption{
Local density of states $N({\mib r},E)$ in the metallic VIC case of 
Fig. \protect\ref{fig:DOS}(b). 
(a) At the center site of the stripe. 
(b) At the nearest neighbor site to the center site. 
(c) At the next nearest neighbor site to the center site. 
(d) At the farthest site from the stripe. 
With leaving the stripe [(a) to (d)], the contribution from the 
mid-gap state disappears. Dashed lines show the up-spin contribution. 
$U=3.6t$, $t'/t=-1/6$, $n_{\rm h}=1/16=0.0625$, $\delta/n_{\rm h}=3/4$. 
The Fermi energy $E_{\rm F}$ is at 0.
}
\label{fig:LDOS}
\end{figure}

This $n_{\rm h}$-dependence of the $m_i$-profile is expected to be observed 
by several experiments. 
We show the $n_{\rm h}$-dependence of the maximum ${\rm max}|m_i|$, and the 
variance $\langle m_i^2 \rangle^{1/2}$. 
As shown in Fig. \ref{fig:mag}, ${\rm max}|m_i|$ sustains a constant 
for  $ 0 < n_{\rm h} < 0.06$,  but $\langle m_i^2 \rangle^{1/2}$ decreases 
gradually with increasing $n_{\rm h}$. 
That is, in the experiments which detect the maximum of the magnetic 
moment such as the neutron scattering, the magnetization does not 
depend on $n_{\rm h}$ for small $n_{\rm h}$. 
But, in the experiments which detect the spatial average of the 
magnetic moment such as NMR, the magnetization gradually decreases 
with increasing $n_{\rm h}$.  
This distinction is observed in Cr.\cite{MachidaCr,Fawcett} 
As seen in Fig. \ref{fig:mag-dis}, the distribution function $P(m)$ 
of the magnetic moment (i.e., the resonance line shape of NMR or 
$\mu$SR) changes its shape reflecting the change from  the square 
wave to the sinusoidal wave. 
At smaller $n_{\rm h}$, $P(m)$ splits into two sharp peaks around $\pm {\rm 
max}|m_i|$ as in the AF state. 
At large $n_{\rm h}$, $P(m)$ distributes broadly between the two peaks. 


As for the profile of the charge density, in the limit of  large 
$n_{\rm h}$, $n_{2{\mibs Q}}/m_{{\mibs Q}}$ is finite and other $n_{l{\mibs Q}} 
\rightarrow 0$ ($l \ge 4$). 
It means that $n_i$ is reduced to the simple sinusoidal wave with 
half period of $m_i$. 
In the limit of small $n_{\rm h}$, $n_{l{\mibs Q}} \sim -(-1)^{l/2} n_{\rm h}  
\rightarrow 0$ ($l \ge 2$: even integer). 
It reflects that the excess carriers are restricted in the 
$\delta$-function-like stripe region whose width is a few sites, and 
that the other wide region is the insulating half-filled AF state.

The spatial profile of the stripe structure obtained here is 
qualitatively the same as that of the insulator state at 
$t'=0$.\cite{Machida,Kato,Zaanen}  
Our results show that the solitonic profile  in the insulator state 
remains unchanged even in the metallic state. 
But the electronic state is changed in the metallic state. 
We discuss it in the following sections.

\begin{fullfigure}[t]
\begin{center}
\leavevmode
\epsfbox{fig11.epsi}
\end{center}
\caption{
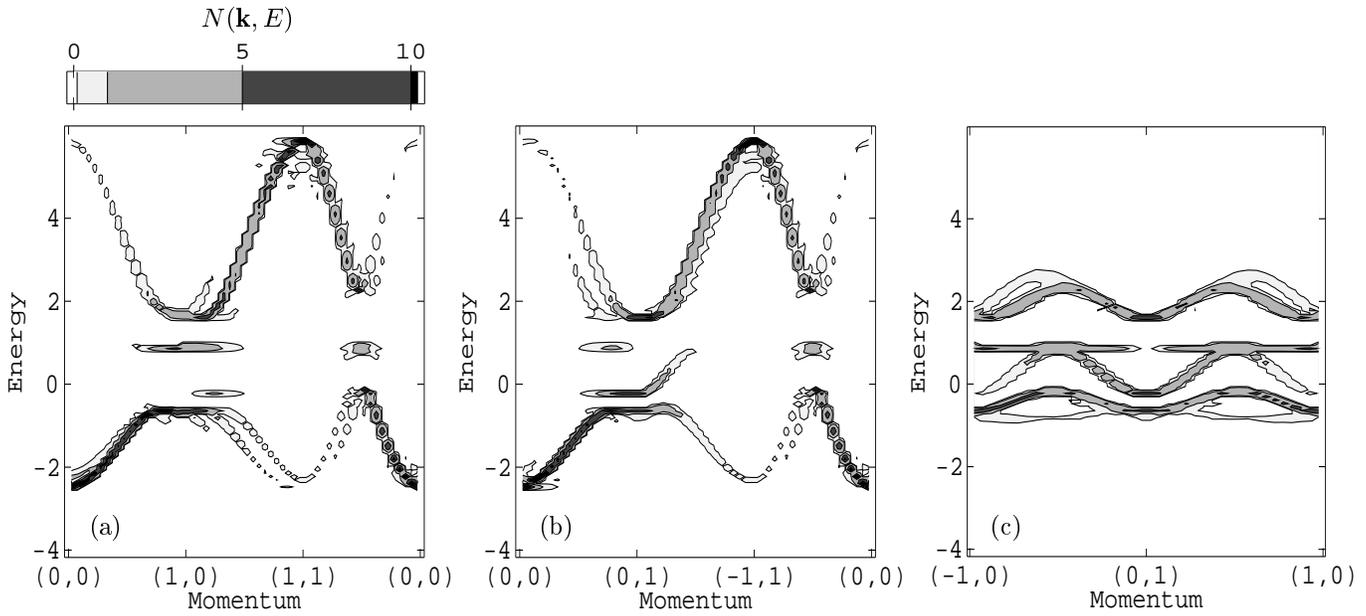
Contour map of the spectral weight $N({\mib k},E)$  along the symmetry 
lines in the original Brillouin zone for the metallic VIC case of 
Fig. \protect\ref{fig:disp-v}(c). 
(a) Along the line $(0,0)$ - $(1,0)$ - $(1,1)$ - $(0,0)$ in 
reciprocal lattice units.  
(b) Along the line $(0,0)$ - $(0,1)$ - $(-1,1)$ - $(0,0)$. 
(c) Along the line $(-1,0)$ - $(0,1)$ - $(1,0)$. 
$U=3.6t$, $t'/t=-1/6$, $n_{\rm h}=1/16=0.0625$, $\delta/n_{\rm h}=3/4$. 
The Fermi energy $E_{\rm F}$ is at $0$. 
Energy $E$ is scaled by $t$. 
}
\label{fig:KDOSv}
\end{fullfigure}
\begin{fullfigure}[t]
\begin{center}
\leavevmode
\epsfxsize=17.0cm  
\epsfbox{fig12s.epsi}
\end{center}
\caption{
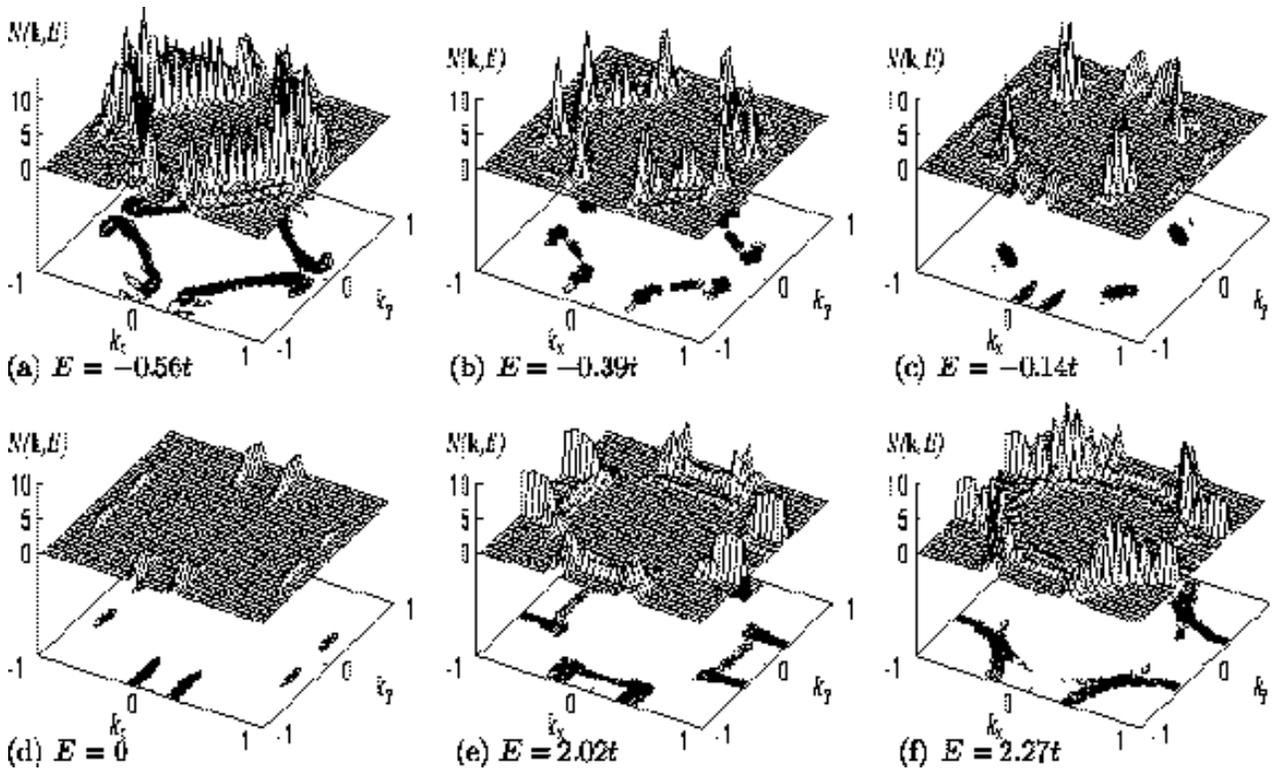
Spectral weight $N({\mib k},E)$ in $2 \pi \times 2 \pi $ ${\mib 
k}$-space at fixed $E$ in the VIC case of Fig. 
\protect\ref{fig:KDOSv}. 
$E/t=-0.56$ (a), $-0.39$ (b), $-0.14$ (c), 0 (d), 2.02 (e), 2.27 (f). 
We see the valence band in (a)-(c), the mid-gap state in (c) and (d), the conduction band in 
(e) and (f).
$k_x$ and $k_y$ are scaled by $\pi$. 
$U=3.6t$, $t'/t=-1/6$, $n_{\rm h}=1/16=0.0625$, $\delta/n_{\rm h}=3/4$. 
The Fermi surface for $\delta/n_{\rm h}=3/4$ corresponds to the ridges in (d). 
The Fermi surface for $\delta/n_{\rm h}=1/2$ has the same structure as 
the ridges in (c). 
}
\label{fig:KDOSev}
\end{fullfigure}
\begin{fullfigure}[t]
\begin{center}
\leavevmode
\epsfbox{fig13.epsi}
\end{center}
\caption{
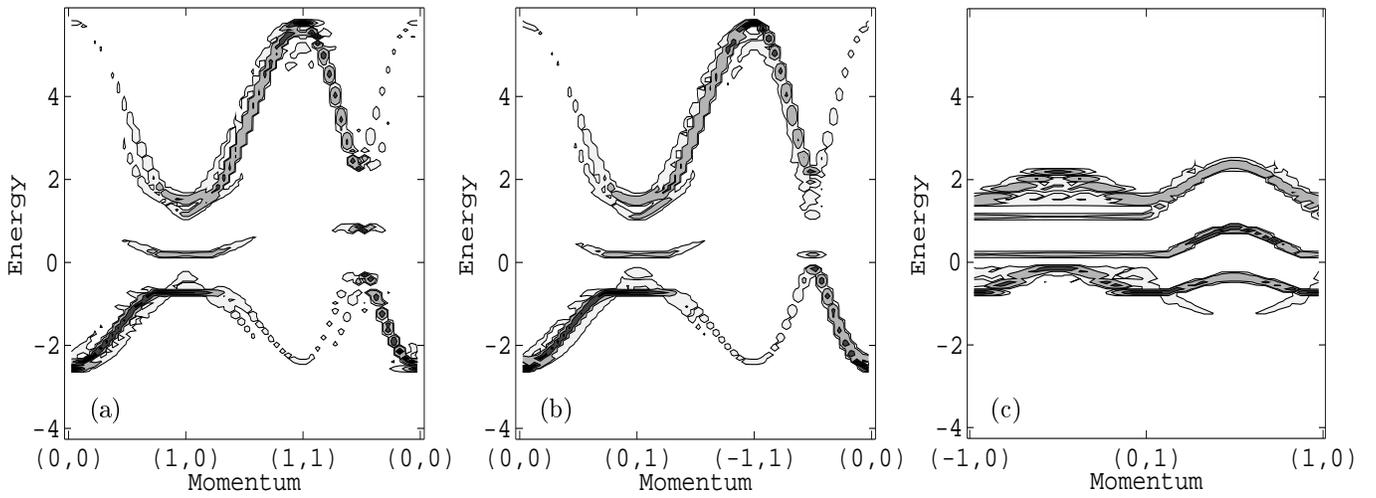
The same as Fig. \protect\ref{fig:KDOSv}, but for the insulator DIC 
case of Fig. \protect\ref{fig:disp-d}(b).
$U=3.6t$, $t'/t=-1/6$, $n_{\rm h}=1/16=0.0625$, $\delta/n_{\rm h}=1/2$.
}
\label{fig:KDOSd}
\end{fullfigure}
\begin{fullfigure}[t]
\begin{center}
\leavevmode
\epsfxsize=17.0cm  
\epsfbox{fig14s.epsi}
\end{center}
\caption{
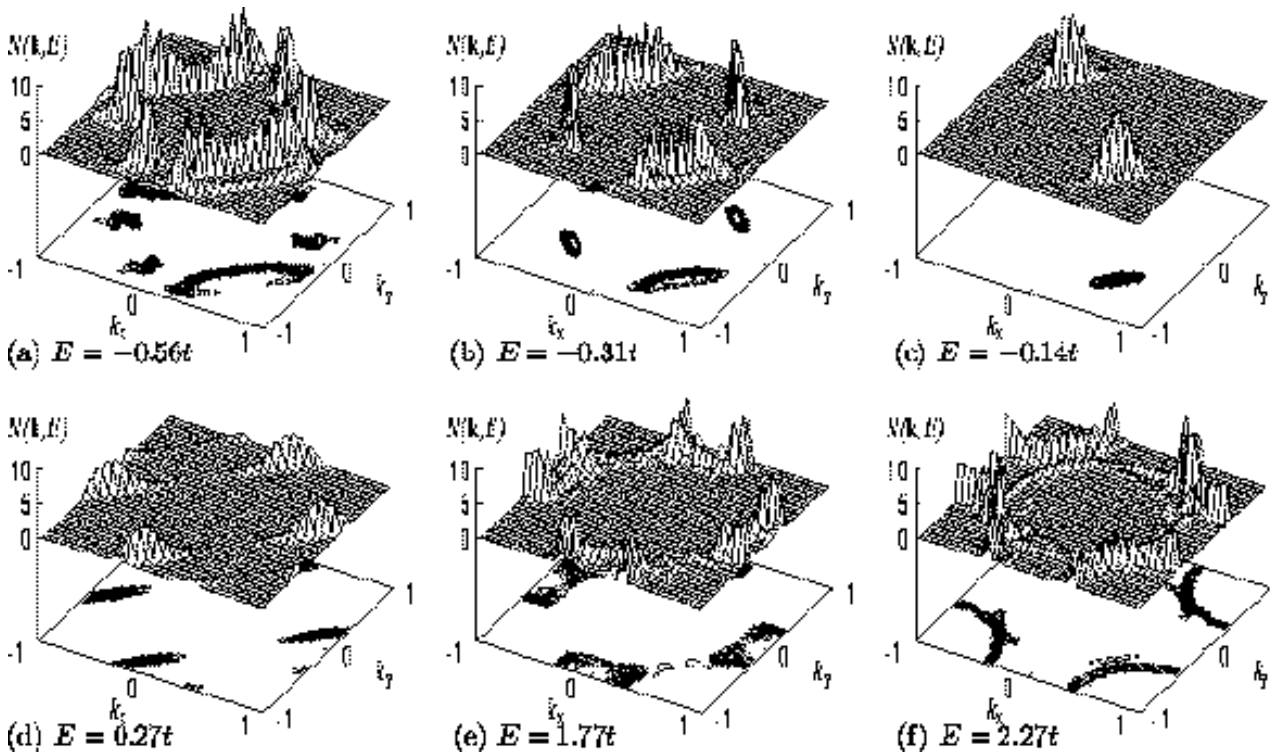
Spectral weight $N({\mib k},E)$ in $2 \pi \times 2 \pi $ 
${\mib k}$-space at fixed $E$ in the DIC case of Fig. 
\protect\ref{fig:KDOSd}.
$E/t=-0.56$ (a), $-0.31$ (b), $-0.14$ (c), 0.27 (d), 1.77 (e), 
2.27 (f). 
We see the valence band in (a)-(c), the mid-gap state in (d), the conduction band in (e) 
and (f). 
$k_x$ and $k_y$ are scaled by $\pi$. 
$U=3.6t$, $t'/t=-1/6$, $n_{\rm h}=1/16=0.0625$, $\delta/n_{\rm h}=1/2$.
}
\label{fig:KDOSed}
\end{fullfigure}

\section{Local Density of States}
\label{sec:LDOS}

In terms of the eigen-energy  and the wave-function,  Green's 
function can be written as 
\begin{equation}
G_\sigma({\mib r},{\mib r}',E)=\sum_{{\mibs k}_0,\alpha}
\frac{u_{{\mibs k}_0,\sigma,\alpha}({\mib r}) 
u^\ast_{{\mibs k}_0,\sigma,\alpha}({\mib r'})}
{E+ {\rm i} \eta - E_{{\mibs k}_0,\sigma,\alpha}}, 
\label{eq4.1}
\end{equation}
where 
\begin{equation} 
u_{{\mibs k}_0,\sigma,\alpha}({\mib r}_i)=N_k^{-1/2} \sum_m 
{\rm e}^{{\rm i}({\mibs k}_0+m{\mibs Q})\cdot {\mibs r}_i} 
u_{{\mibs k}_0,\sigma,\alpha,m}  
\label{eq:4.2} 
\end{equation}
and $\eta(>0)$ is an infinitesimally small constant. 
We obtain the LDOS by 
\begin{eqnarray}
N_\sigma({\mib r}_i,E)&=&
-\frac{1}{\pi}{\rm Im} G_\sigma({\mib r}_i,{\mib r}_i,E) 
\nonumber \\ 
&=&
\sum_{{\mibs k}_0,\alpha}|u_{{\mibs k}_0,\sigma,\alpha}({\mib r}_i)|^2 
\delta(E-E_{{\mibs k}_0,\sigma,\alpha}).
\label{eq:4.3}
\end{eqnarray}
The DOS is the spatial average of LDOS, i.e., 
\begin{equation}
N_\sigma(E)=N_k^{-1}\sum_{i}N_\sigma({\mib r}_i,E) 
=N_k^{-1} \sum_{{\mibs k}_0,\alpha}\delta(E-E_{{\mibs 
k}_0,\sigma,\alpha}).
\label{eq:4.4}
\end{equation}

Figure \ref{fig:DOS} shows the DOS for DIC and VIC in the case 
$n_{\rm h}=1/16=0.0625$. 
The mid-gap state is seen within the large AF gap. 
In the insulator DIC case, there is a small gap between the mid-gap 
states and the valence band, and $E_{\rm F}$ is located in the small gap. 
In the metallic VIC case, the small gap is buried in this DOS plot. 
For the case $\delta =\frac{1}{2}n_{\rm h}$, $E_{\rm F}(=0)$ is 
located at the overlapped mid-gap state and the valence band. 
With increasing $\delta$ at fixed $n_{\rm h}$, $E_{\rm F}$ is shifted to 
higher energy without qualitative change of the DOS structure. 
But, in the large $\delta(\sim n_{\rm h})$ case for 
large $n_{\rm h}(\ge 0.1)$, 
the mid-gap state and the valence band  are separated in the DOS plot, where 
$E_{\rm F}$ is located in the mid-gap states. 
As a result, even in the metallic state, a gap feature in the 
electron state still remains. 
This point is discussed further in the next section. 


Next, we consider the LDOS, $N({\mib r}_i,E)$. 
It is shown in Fig. \ref{fig:LDOS}. 
The contribution of the mid-gap state is enhanced at the center of 
the stripe, and  decreased with leaving the center. 
Far from the stripe, the LDOS is like an AF-gapped spectrum since the 
mid-gap state's contribution vanishes. 
For $\delta=\frac{3}{4}n_{\rm h}$ (and also for $\delta=n_{\rm h}$), 
$E_{\rm F}$ located in the mid-gap state is in the AF gap outside the stripe 
region [Fig. \ref{fig:LDOS}(d)]. 
Then only the stripe region is metallic, and the other AF region is 
an insulator. 
If $\delta = \frac{1}{2}n_{\rm h}$ in the metallic VIC case, 
where $E_{\rm F}$ is effectively shifted to lower energy, 
$E_{\rm F}$ touches the mid-gap state in the stripe region 
and the valence band in the outside region. 
Then, both the stripe region and outer AF region are metallic. 
If these characteristic features of the ${\mib r}$-resolved DOS 
structure are observed by experiments such as STM, it may be a 
powerful means to confirm the solitonic features of the stripe 
structure. 


\section{Spectral Weight}
\label{sec:KDOS}

In this section, we study the ${\mib k}$-resolved DOS, i.e., spectral 
weight at each ${\mib k}$ point. 
It is given by 
\begin{eqnarray} &&
N_\sigma({\mib k},E)=
-\frac{1}{\pi}{\rm Im} G_\sigma({\mib k},E) 
\nonumber \\ &&
= \sum_{{\mibs k}_0,\alpha,m}|u_{{\mibs k}_0,\sigma,\alpha,m}|^2 
\delta(E-E_{{\mibs k}_0,\sigma,\alpha})\delta({\mib k}_0+m{\mib Q}-{\mib 
k}), 
\nonumber \\ &&
\label{eq:5.1}
\end{eqnarray}
where 
\begin{equation}
G_\sigma({\mib k},E)=N_k^{-1}\sum_{{\mibs r},{\mibs r}'}
{\rm e}^{-{\rm i} {\mibs k}\cdot({\mibs r}-{\mibs r}')} 
G_\sigma({\mib r},{\mib r}',E). 
\label{eq:5.2}
\end{equation}
In eq. (\ref{eq:5.1}), the momentum of the center of mass coordinate 
$({\mib r}+{\mib r}')/2$ is set to be 0, since we consider the spatial 
average.

First, we consider the VIC case of Fig. \ref{fig:disp-v}(c) for 
$n_{\rm h}=1/16=0.0625$ and $\delta=\frac{3}{4}n_{\rm h}$. 
Figures \ref{fig:KDOSv} shows the contour map of $N({\mib 
k},E)=\sum_{\sigma}N_\sigma({\mib k},E)$ along the symmetry lines in 
the original Brillouin zone. 
We also calculate $N({\mib k},E)$ for other $n_{\rm h}$ cases, and obtain 
qualitatively the same structure. 
While the energy dispersion splits into $N$-bands by the $N$-site 
periodicity, the spectral weight in $2 \pi \times 2 \pi$ ${\mib 
k}$-space has the structure reflecting the original dispersion 
$\epsilon({\mib k})$ as shown in Fig. \ref{fig:KDOSv}, since the 
amplitude of the wave function $u_{{\mibs k}_0,\sigma,\alpha,m}$ for 
each $m$ reflects the original dispersion $\epsilon({\mib k})$. 
Some $m{\mib Q}$-shifted $\epsilon({\mib k})$ by the band folding also 
appear in the figure.  
While their amplitude are usually small, they form the SDW gap 
and the mid-gap states.  
Darker region in Fig. \ref{fig:KDOSv}  means larger $N({\mib k},E)$. 
As the stripe is a 1D structure, $N({\mib k},E)$ is two-fold 
symmetric.  
Then, there are differences of $N({\mib k},E)$ between $(1,0)$ and 
$(0,1)$ directions in the VIC case. 
In Fig. \ref{fig:KDOSv}, $N({\mib k},E=E_{\rm F})$ has little 
amplitude around $(1,0)$. 
But near $(0,1)$, the mid-gap state appears with large amplitude at 
$E_{\rm F}$. 
There, we can see the split of the mid-gap state and the valence band below 
$E_{\rm F}$. 
Around $(\frac{1}{2},\frac{1}{2})$ and $(-\frac{1}{2},\frac{1}{2})$, 
we see a gapped structure, where the mid-gap state does not appear at 
$E_{\rm F}$.


These features of the spectral weight are consistent with those 
obtained by the ARPES experiments on LSCO.~\cite{Ino1,Ino2,Ino3} 
Since there are two-type domains of the $x$-direction and 
the $y$-direction stripe in the material, we observe the spectral weight 
as $N_{\rm D}({\mib k},E)=[N({\mib k},E)+N(k_x \leftrightarrow 
k_y,E)]/2$, i.e., the overlap of $N({\mib k},E)$ in Figs. 
\ref{fig:KDOSv}(a) and \ref{fig:KDOSv}(b). 

When $\delta$ is shifted keeping $n_{\rm h}$ fixed, we confirm that $N({\mib 
k},E)$ is not qualitatively changed except for the position of 
$E_{\rm F}$.  
With increasing $\delta$, $E_{\rm F}$ is shifted to upper energy. 
For larger $\delta$ such as $\delta=n_{\rm h}$, the gap edge at 
$(\frac{1}{2},\frac{1}{2})$ is located at deeper position below 
$E_{\rm F}$. 
The ARPES experiments~\cite{Ino1,Ino2,Ino3} reported that the gap at 
$(\frac{1}{2},\frac{1}{2})$ for $x=0.15$ disappears with increasing 
$n_{\rm h}$. 
This gap disappearance can be explained as follows in our scenario. 
In the elastic neutron scattering data,~\cite{Wakimoto,Wakimoto2,Suzuki,TranquadaR}  
with increasing $n_{\rm h}$, $\delta$ shows saturation and it is shifted 
from $\delta \sim n_{\rm h}$ to $\delta \sim \frac{1}{2}n_{\rm h}$. 
When $\delta$ approaches $\frac{1}{2}n_{\rm h}$, $E_{\rm F}$ is effectively 
shifted toward lower energy, and touches the top of the valence band near 
$(\frac{1}{2},\frac{1}{2})$. 
Then, the  gap in the $(\frac{1}{2},\frac{1}{2})$-direction disappears. 
As $E_{\rm F}$ approaches the gap-edge at  
$(\frac{1}{2},\frac{1}{2})$ gradually, the gap disappearance is 
``observed'' at larger $\delta$ than $\delta=\frac{1}{2}n_{\rm h}$, if the 
energy resolution is broad in the experiment. 

To clearly see the ${\mib k}$-dependent SDW gap, we consider $N({\mib 
k},E)$ in $2 \pi \times 2 \pi$ ${\mib k}$-space at fixed $E$. 
It is shown in Fig. \ref{fig:KDOSev} for VIC. 
From (a) to (f), $E$ is increased. 
In Figs. \ref{fig:KDOSev}(a)-(c), the valence band appears. 
Its behavior is understood by considering $E_-({\mib k})$ of the C-AF 
case in eq. (\ref{eq:3.1}). 
Below $E_{\rm F}$, $N({\mib k},E)$ has ridges along the contour line 
for $E=E_-({\mib k})$. 
Far from $E_{\rm F}$, it reflects the original dispersion  
$\epsilon({\mib k})$ [Fig. \ref{fig:KDOSev}(a)].
For $E_-(\pi(1,0)) < E < E_-(\pi(\frac{1}{2},\frac{1}{2}))$, $N({\mib 
k},E)$ disappears near $(1,0)$ and $(0,1)$, because the AF gap opens 
at these ${\mib k}$-points. 
With increasing $E$, the remained contour lines of $E_-({\mib k})$ 
decrease their length, and converge upon $(\frac{1}{2},\frac{1}{2})$ 
[Figs. \ref{fig:KDOSev}(b) and \ref{fig:KDOSev}(c)]. 
Near $E_{\rm F}$, we also see the fragmentation of the ridges due to 
the $x$-direction 1D stripe structure [Figs. \ref{fig:KDOSev}(b)].  
The valence band contribution disappears for 
$E>E_-(\pi(\frac{1}{2},\frac{1}{2}))$. 


The mid-gap state appears in Fig. \ref{fig:KDOSev}(c) and 
\ref{fig:KDOSev}(d). 
On raising $E$, it first appears near $(0,1)$ [and (1,0) with small 
amplitude] at the bottom of the mid-gap bands in addition to the valence band 
contribution [Figs. \ref{fig:KDOSev}(c)]. 
The ``Fermi surface'' in the VIC case corresponds to the ridges shown 
in Fig. \ref{fig:KDOSev}(d), where the valence band contribution already 
disappears. 
Since the mid-gap state is localized along the $x$-directional 
stripe, $N({\mib k},E)$ is distributed on lines perpendicular to $k_x$ 
direction, like a 1D Fermi surface. 
As it also reflects the original Fermi surface of $\epsilon({\mib 
k})$, the amplitude of $N({\mib k},E)$ is large near $(0,1)$. 
Then in the spectral weight of the  domain structure $N_{\rm D}({\mib 
k},E)$, the Fermi surface exists near $(0,1)$ and $(1,0)$, but 
disappears near $(\frac{1}{2},\frac{1}{2})$. 
This ${\mib k}$-dependent Fermi surface structure is consistent with 
the ARPES results.~\cite{Ino2,Ino3}  
We obtain qualitatively the same Fermi surface also for $\delta=n_{\rm h}$. 
It is qualitatively the same results of the case  $t'=0$ under the 
test stripe potential in ref. ~\citen{Salkola}.  
For $\delta=\frac{1}{2}n_{\rm h}$, since $E_{\rm F}$ is decreased 
effectively, the Fermi surface has the same structure as shown in 
Fig. \ref{fig:KDOSev}(c), i.e., the valence band contribution also 
appears near $(\frac{1}{2},\frac{1}{2})$ in addition to  the mid-gap state's 
contribution near $(1,0)$. 
This feature is contrasted with the $\delta=\frac{3}{4}n_{\rm h}$ case as 
mentioned above. 

With increasing $E$, the larger ridge-wall near $(0,1)$ shown in Fig. 
\ref{fig:KDOSev}(d) goes away from the line $k_x=0$, and the small 
ridge-wall near $(1,0)$ approaches $k_x=0$. 
In this shift, the ridge of $N({\mib k},E)$ has large peaks near the 
${\mib k}$-points of the original Fermi surface by $\epsilon({\mib 
k})$. 
Then, if $\delta$ is randomly distributed (it means effectively that 
$E_{\rm F}$ is randomly distributed), the original Fermi surface 
shape of $\epsilon({\mib k})$ appears as an average.~\cite{Salkola} 

The conduction band contribution appears in Figs. \ref{fig:KDOSev}(e) and 
\ref{fig:KDOSev}(f). 
It qualitatively follows $E_+({\mib k})$ dispersion in eq. 
(\ref{eq:3.1}). 
For $E_+(\pi(1,0))<E<E_+(\pi(\frac{1}{2},\frac{1}{2}))$, the conduction band 
contribution appears only near $(1,0)$ and $(0,1)$ [Fig. 
\ref{fig:KDOSev}(e)].  
With increasing $E$, it spreads and approaches 
$(\frac{1}{2},\frac{1}{2})$, and it is connected each other. 
For $E>E_+(\pi(\frac{1}{2},\frac{1}{2}))$, the ridges of $N({\mib 
k},E)$  reflect the original dispersion $\epsilon({\mib k})$. 
We also see the small amplitude of the circle ridge around $(0,0)$, 
which is the ${\mib Q}$-shift of the original $\epsilon({\mib k})$ 
contribution.

Next, we consider the DIC case calculated for the same parameter set. 
It is shown in Figs. \ref{fig:KDOSd} and \ref{fig:KDOSed}. 
There, $l{\mib Q}$-shifted $\epsilon({\mib k})$ dispersion curves form 
the valence band, the conduction band and the mid-gap state 
as in the VIC case. 
As shown in Fig. \ref{fig:KDOSd}, a small gap opens between the valence band 
and the mid-gap state in the DIC case, and $E_{\rm F}$ is in the gap. 
Reflecting the diagonal-direction stripe, $N({\mib k},E)$ is two-fold 
symmetric. 
The differences appear between the 
$(\frac{1}{2},\frac{1}{2})$-direction and the 
$(-\frac{1}{2},\frac{1}{2})$-direction. 
In Fig. \ref{fig:KDOSd}(c), $N({\mib k},E)$ shows flat dispersion near 
$E_{\rm F}$ along the line $(-1,0)$ - $(0,1)$. 
It reflects the 1D-like dispersion due to the stripe, as discussed 
below. 


To clearly see the ${\mib k}$-dependent SDW gap for DIC, we consider 
$N({\mib k},E)$ at fixed $E$ as shown in Fig. \ref{fig:KDOSed}. 
The valence band contribution appears in Figs. \ref{fig:KDOSed}(a)-(c). 
For $E_-(\pi (1,0))<E<E_-(\pi(\frac{1}{2},\frac{1}{2}))$, $N({\mib 
k},E)$ near $(1,0)$ and $(0,1)$ disappears as in the VIC case. 
Further, by the effect of the diagonal stripe, the remained 
ridge-wall near $(\frac{1}{2},\frac{1}{2})$ is split into two peaks 
[Fig. \ref{fig:KDOSed}(a)], and shifted toward  
$(\frac{1}{2},\frac{1}{2})$ [Fig. \ref{fig:KDOSed}(b)], and 
disappears [Fig. \ref{fig:KDOSed}(c)], with increasing $E$. 
As for the ridge-wall near $(-\frac{1}{2},\frac{1}{2})$, its length 
monotonically shrinks on raising $E$ [Figs. \ref{fig:KDOSed}(a) and 
\ref{fig:KDOSed}(b)], but the ridge survives for a while after the 
$(\frac{1}{2},\frac{1}{2})$-ridge disappears 
[Fig. \ref{fig:KDOSed}(c)]. 

Figure \ref{fig:KDOSed}(d) shows the mid-gap state's contribution. 
As the mid-gap state is localized along the diagonal stripe, its 
dispersion is 1D-like. 
The ridge-wall approaches the line $k_x=k_y$ with increasing $E$. 
The ridge-wall has two large peaks. 
Their peak-positions are near the contour line of the original 
dispersion $\epsilon({\mib k})$. 

The conduction band contribution appears in Figs. \ref{fig:KDOSed}(e) and 
\ref{fig:KDOSed}(f). 
At the bottom of the conduction band, the peaks of $N({\mib k},E)$ appear around 
$(1,0)$ and $(0,1)$ [Fig. \ref{fig:KDOSed}(e)]. 
And the additional ridges connect these peaks along the line $(1,0)$ 
- $(0,-1)$. 
The peaks are separated along the $(1,0)$ - $(0,1)$ direction. 
This 1D-like ridge comes from the diagonal stripe structure. 
With increasing $E$, the contributions near $(1,0)$ and $(0,1)$ 
spread out, and they are connected each other. 
Then, $N({\mib k},E)$ has a connected ridge along the contour line of 
the original dispersion $\epsilon({\mib k})$ 
[Fig. \ref{fig:KDOSed}(f)].  
We also see the ${\mib Q}$-shifted contribution of $\epsilon({\mib k})$ 
as small ridges. 

\section{Optical Conductivity}
\label{sec:optical}

We calculate the optical conductivity by the inter-band 
vertical transition within the reduced Brillouin zone. 
Since there are $N$ bands in the stripe structure with the $N$-site 
periodicity as shown in Figs. \ref{fig:disp-d} and \ref{fig:disp-v}, 
the inter-band absorption occurs by these $N$ bands. 
The optical conductivity ${\rm Re}\sigma_{j,j'}(\omega)$ $(j,j'=x,y)$ 
is given by the Kubo-Greenwood 
formula.~\cite{Harrison,Kubo,Greenwood} 
In our notation, it is written as 
\begin{eqnarray} &&
{\rm Re}\sigma_{j,j'}(\omega)
\nonumber \\ && 
=\frac{\pi }{\omega N_k} 
\sum_{{\mibs k}_0, \alpha, \alpha',\sigma}
f(E_{{\mibs k}_0,\alpha,\sigma})
[1-f(E_{{\mibs k}_0,\alpha',\sigma})] 
\nonumber \\ && \times
\langle {\mib k}_0,\alpha,\sigma |J_j|{\mib k}_0,\alpha',\sigma \rangle 
\langle {\mib k}_0,\alpha,\sigma |J_{j'}|{\mib k}_0,\alpha',\sigma 
\rangle 
\nonumber \\ && \times
[\delta(E_{{\mibs k}_0,\alpha,\sigma} - E_{{\mibs k}_0,\alpha',\sigma} + 
\omega)
-\delta(E_{{\mibs k}_0,\alpha,\sigma} - E_{{\mibs k}_0,\alpha',\sigma} - 
\omega)].
\nonumber \\
\label{eq:6.1}
\end{eqnarray}
Equation (\ref{eq:6.1}) is derived from the current-current correlation 
function. 
The $x$-component of the current operator is defined as 
$\hat{J}_x({\mib r}_i)=-{\rm i}|e| \sum_\sigma \{ 
t(a_{i+\hat{x},\sigma}^\dagger - a_{i-\hat{x},\sigma}^\dagger) 
+t' ( a_{i+\hat{x}+\hat{y},\sigma}^\dagger 
     +a_{i+\hat{x}-\hat{y},\sigma}^\dagger
     -a_{i-\hat{x}+\hat{y},\sigma}^\dagger
     -a_{i-\hat{x}-\hat{y},\sigma}^\dagger )\} a_{i,\sigma}  $ 
in our square lattice 
case, where $i+\hat{x}$ and $i+\hat{y}$ mean  $(i_x+1,i_y)$ and 
$(i_x,i_y+1)$, respectively. 
Then, the $x$-component of the oscillator strength 
in eq. (\ref{eq:6.1}) is given by 
\begin{eqnarray} &&
\langle {\mib k}_0,\alpha,\sigma |J_x|{\mib k}_0,\alpha',\sigma \rangle 
=|e| \sum_i [ t \{ u_{{\mibs k}_0,\alpha ,\sigma} ({\mibs r}_{i+\hat{x}})   
\nonumber \\ &&
                  -u_{{\mibs k}_0,\alpha ,\sigma} ({\mibs r}_{i-\hat{x}})  \}
+t' \{ u_{{\mibs k}_0,\alpha ,\sigma} ({\mibs r}_{i+\hat{x}+\hat{y}})
      +u_{{\mibs k}_0,\alpha ,\sigma} ({\mibs r}_{i+\hat{x}-\hat{y}})
\nonumber \\ &&
      -u_{{\mibs k}_0,\alpha ,\sigma} ({\mibs r}_{i-\hat{x}+\hat{y}})
      -u_{{\mibs k}_0,\alpha ,\sigma} ({\mibs r}_{i-\hat{x}-\hat{y}})
 \} ] u_{{\mibs k}_0,\alpha',\sigma} ({\mibs r}_i) 
\nonumber \\ &&
\label{eq:6.2}
\end{eqnarray}
instead of the usual form $\langle \psi_{\mibs k} | \nabla_x | 
\psi_{{\mibs k}'} \rangle $. 

\begin{figure}[t]
\begin{center}
\leavevmode
\epsfbox{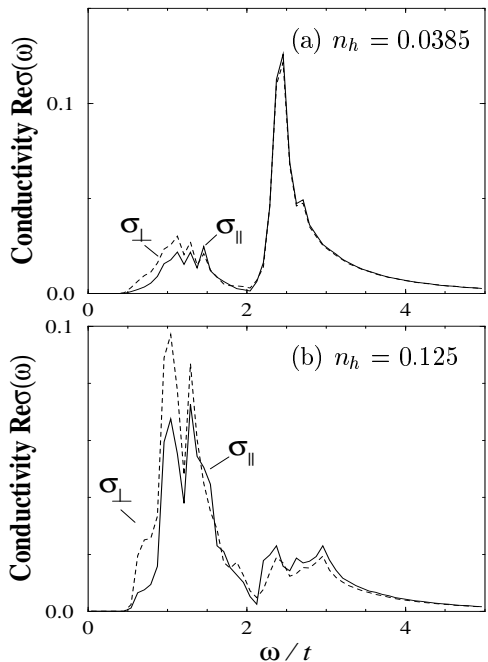}
\end{center}
\caption{
Optical conductively ${\rm Re}\sigma_\parallel(\omega)$ [solid lines] 
and ${\rm Re}\sigma_\perp(\omega)$ [dashed lines] for VIC 
in an arbitrary unit. 
${\rm Re}\sigma_\parallel(\omega)$ (${\rm Re}\sigma_\perp(\omega)$) 
is for the field parallel (perpendicular) to the stripe.
The Drude absorption is not included in this figure. 
(a) $n_{\rm h}=1/26=0.0385$, $\delta/n_{\rm h}=13/18$. 
(b) $n_{\rm h}=1/8=0.125$, $\delta/n_{\rm h}=5/7$. 
$U=3.6t$, $t'/t=-1/6$. 
}
\label{fig:opcv}
\end{figure}

Figure \ref{fig:opcv} shows ${\rm Re}\sigma_\parallel (\omega)$ for 
the parallel field to the stripe and ${\rm Re}\sigma_\perp (\omega)$ 
for the perpendicular field in the VIC case.  
Here, we do not consider the Drude absorption by the 
intra-band inelastic scattering. 
For $n_{\rm h}=1/26=0.0385$ [Fig. \ref{fig:opcv}(a)], there is a large peak 
near $\omega \sim 2.5t$. 
It comes from the valence band $\rightarrow$ conduction band absorption with larger $\omega$ 
than the AF gap, and survives also in the C-AF state. 
We call the peak as the ``AF peak''. 
The peaks at $\omega/t=1 \sim 1.5$ comes from the absorption related 
to the mid-gap states.  
We call them as the ``mid-gap peaks''. 
They vanish in the C-AF state. 
For larger $n_{\rm h}$ [Fig. \ref{fig:opcv}(b)], the AF peak is suppressed 
and the mid-gap peaks are enhanced. 
The mid-gap peaks of ${\rm Re}\sigma_\perp (\omega)$ are slightly 
shifted to lower $\omega$ compared with ${\rm Re}\sigma_\parallel 
(\omega)$. 

\begin{figure}[t]
\begin{center}
\leavevmode
\epsfbox{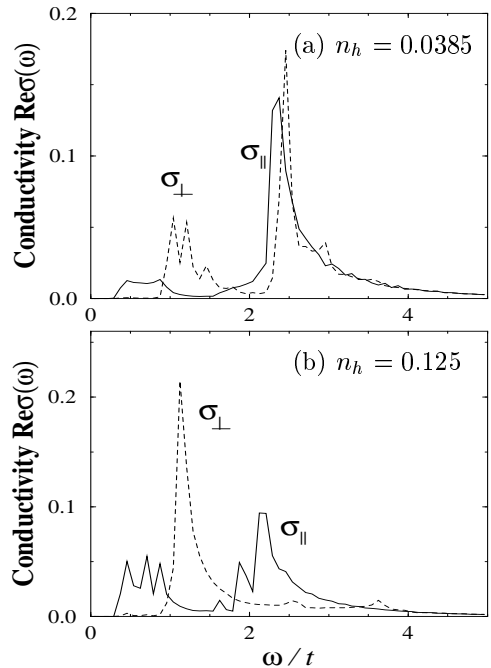}
\end{center}
\caption{
Optical conductively ${\rm Re}\sigma_\parallel(\omega)$ [solid lines] 
and ${\rm Re}\sigma_\perp(\omega)$ [dashed lines] for DIC 
in an arbitrary unit. 
(a) $n_{\rm h}=1/26=0.0385$.
(b) $n_{\rm h}=1/8=0.125$.
$U=3.6t$, $t'/t=-1/6$, $\delta/n_{\rm h}=1/2$.
}
\label{fig:opcd}
\end{figure}

The DIC case is shown in Fig. \ref{fig:opcd}. 
With increasing $n_{\rm h}$, the AF peak is suppressed and the mid-gap 
peaks are enhanced, as in the VIC case. 
There are large differences between ${\rm Re}\sigma_\parallel 
(\omega)$ and ${\rm Re}\sigma_\perp (\omega)$ in the DIC case. 
The mid-gap peaks of ${\rm Re}\sigma_\perp (\omega)$ are large, and 
they are located at larger $\omega$.  
For $n_{\rm h}=1/8=0.125$, the AF peak vanishes for ${\rm Re}\sigma_\perp 
(\omega)$, but still has large height for ${\rm Re}\sigma_\parallel 
(\omega)$. 

The experiments of the optical conductivity were performed on LSCO 
[ref. \citen{Uchida}] and La$_{2-x}$Sr$_x$NiO$_4$ [ref. \citen{Ido}].
There, the broad peak develops in the mid-infrared region 
with increasing $n_{\rm h}$. 
It may be related to the mid-gap peak's contribution of the stripe state. 

\section{Summary and Discussions}
\label{sec:summary}

The stripe structure in high-$T_{\rm c}$ cuprates and its electronic 
structure are studied by the self-consistent mean-field theory of the 
Hubbard model. 
The SDW gapped insulator is changed to a metal by introducing the 
realistic Fermi surface topology in VIC. 
For appropriate $U$, DIC is changed to VIC at critical $n_{\rm h}$ with 
increasing $n_{\rm h}$.  
The DIC is insulator, and VIC can be metallic. 
In the insulator, $\delta = \frac{1}{2}n_{\rm h}$. 
But, in the metallic VIC, $\delta$ can be larger than $\frac{1}{2} n_{\rm h}$. 
These features seem to be consistent with the experimental results of 
the elastic neutron 
scattering,~\cite{Wakimoto,Wakimoto2,Suzuki,TranquadaR,TranquadaNd,TranquadaNi,Yoshizawa} 
as discussed in \S \ref{sec:profile}. 

The SDW and CDW of the stripe structure are a sinusoidal wave at 
large $n_{\rm h}$. 
They become, respectively, square and solitonic waves at small $n_{\rm h}$. 
There, excess carriers are localized in the stripe region and the 
outer region is half-filled AF. 
The localized state forms the mid-gap state inside the AF gap. 
We calculate the LDOS, the spectral weight and the optical 
conductivity, and clarify the contribution of the mid-gap state. 
In the metallic VIC case, only the mid-gap state appears at $E_{\rm 
F}$ for $\delta/n_{\rm h} \sim \frac{3}{4} $ or larger. 
Then, while the stripe region is metallic locally in space, the outer 
AF region is an insulator. 
Metallicity attains locally on the stripe region for finite $n_{\rm h}$. 
In the spectral weight, the contribution of the mid-gap state has 
large amplitude near $(0,1)$ and little amplitude near 
$(\frac{1}{2},\frac{1}{2})$ at $E_{\rm F}$ in VIC. 
Then, the ``Fermi surface'' disappears near 
$(\frac{1}{2},\frac{1}{2})$, leaving the Fermi surface ``arcs''. 
On the other hand, for $\delta/n_{\rm h} \sim \frac{1}{2}$, both the 
mid-gap state and the  valence band appear at $E_{\rm F}$. 
In this case, both the inside and outside spatial regions of the 
stripe are metallic. 
In the spectral weight, the valence band contribution appears near 
$(\frac{1}{2},\frac{1}{2})$ in addition to the mid-gap state's 
contribution near $(0,1)$. 
Then, the Fermi surface is observed as a connected one. 
These features of the spectral weight are consistent with the ARPES 
experiments,~\cite{Ino1,Ino2,Ino3} as discussed in \S 
\ref{sec:KDOS}.  
As a result, the electronic structure near $E_{\rm F}$ is determined 
by the ratio of the incommensurability and the hole density, $\delta 
/ n_{\rm h}$, which is a key parameter of the problem. 
We expect that the features of the electronic state obtained here are 
examined by the experiments such as  ARPES, STM and the optical 
conductivity. 
These experiments may be powerful methods to obtain vital information 
about the mechanism of the stripe structure in high-$T_{\rm c}$ cuprates. 

So far, the metallic stripe state with $\delta/n_{\rm h} \sim 1$ could not 
be reproduced by the mean field theory, which gives an insulator with 
$\delta/n_{\rm h} =\frac{1}{2}$. 
Then, the stripe state was said to be a anomalous state where the mean 
field theory can not be applied. 
However, by introducing the realistic Fermi surface topology, 
the mean field theory can gives the metallic state with 
$\delta/n_{\rm h} = 0.75 \sim 0.88$ (which approaches 1 depending on 
the Fermi surface topology) even in a simple Hubbard model. 
While the mean field theory over-estimates the stable parameter 
region of the ordered states or the AF moments, the obtained 
features are qualitatively valid as a first approximation 
within the ordered state. 
Then, the mean-field theory can be a first step for further studies 
to consider the effects of the extra interactions or fluctuations, 
and to obtain the behavior of the various physical quantities 
in the stripe state.



\end{document}